\documentclass[reprint,prl]{revtex4-1}
\usepackage{graphicx}
\usepackage{color}
\usepackage{bm}
\usepackage{amsmath}

\begin{document}

\title{Parametric  Dependence of Bound States in the
    Continuum on Periodic Structures}

\author{Lijun Yuan}
\email{ljyuan@ctbu.edu.cn}
\affiliation{College of Mathematics and Statistics, Chongqing Technology and Business University, Chongqing, 
China}

\author{Ya Yan Lu}
\email{mayylu@cityu.edu.hk}
\affiliation{Department of Mathematics, City University of Hong Kong, 
  Kowloon, Hong Kong, China} 
\date{\today}

\begin{abstract}
Bound states in the continuum (BICs) have some unusual properties and
important applications in photonics. A periodic structure sandwiched  
between two homogeneous media is the most popular platform for
observing BICs and realizing their applications. Existing studies  on 
BICs assume the periodic structure has a $C_2$ rotational symmetry about the 
axis perpendicular to the periodic layer. It is known that all BICs 
turn to resonant states with finite quality factors if the periodic
structure is perturbed by a generic perturbation breaking the $C_2$
symmetry, and a typical BIC continues to exist if the perturbation
keeps the $C_2$ symmetry.
We study how typical BICs depend on generic
structural parameters. For a class of BICs with one opening
radiation channel, we show that 
in the  plane of two generic parameters, the BICs exist 
continuously as a curve. 
Consequently, BICs can exist on periodic
structures without the $C_2$ symmetry, and they can be found by tuning
a single structural parameter.
The result is established analytically 
by a perturbation theory with two independent perturbations and
validated by numerical examples. Our study reveals a much
larger family for BICs 
on periodic structures, and provides new opportunities for future
applications. 
\end{abstract}
\maketitle 


A bound state in the continuum (BIC) is a localized eigenmode that
does not couple with compatible waves propagating to or from infinity
\cite{neumann29,hsu16,kosh19}. For photonic systems, BICs can exist in
various structures including waveguides with local distortions
\cite{bulg08}, waveguides with 
lateral leaky channels \cite{plot11,weim13,byk20}, layered anisotropic 
media \cite{gomis17}, rotationally symmetric periodic structures 
\cite{bulg17prl,sadbel19}, and periodic structures sandwiched between
two homogeneous media
\cite{mari08,lee12,hsu13_2,yang14,zhen14,yoon15,gan16,bulg17pra,doel18,zhang18,jin19,huyuan20}. 
Since a BIC can be regarded as a resonant state
with an infinite quality factor ($Q$ factor), it is possible to 
realize resonant states with extremely large $Q$ factors by perturbing the
structures \cite{kosh18,hu18,lijun20,yin20} or varying the
Bloch wavevector \cite{lijun18pra,zhen20b}. Although true BICs can only exist on 
structures that are unbounded in at least one spatial direction, they
have inspired the design of wavelength-scale structures supporting high-$Q$
resonances \cite{rybin17,taghi17}. Due to their unusual properties, 
the BICs are useful in a number of applications such as lasing
\cite{kodi17}, sensing \cite{romano19,yesi19}, filtering and
switching, and can be used to enhance various emmisive  
processes and nonlinear optical effects \cite{lijun17pra,lijun19siam,kosh20}.

On a periodic structure (with one or two periodic directions)
sandwiched between two homogeneous 
media, a BIC is either a standing 
wave or a propagating Bloch mode. In all existing studies
\cite{mari08,lee12,hsu13_2,yang14,zhen14,yoon15,gan16,bulg17pra,doel18,zhang18,jin19},  
the periodic structure is assumed to have at least the $180^\circ$ (i.e. $C_2$) rotational symmetry
about the axis $z$ perpendicular to the periodic layer. 
Some standing waves are symmetry-protected BICs. Their existence can
be rigorously proved, and they are robust against small structural 
perturbations that preserve the $C_2$ symmetry. 
Most propagating BICs are
found on periodic structures with not only the $C_2$ symmetry but also
a reflection symmetry in $z$
\cite{mari08,hsu13_2,yang14,zhen14,gan16,bulg17pra,doel18,zhang18,jin19}.  
The existence and robustness of the propagating 
BICs (as well as the standing waves) have been studied using
topological properties \cite{zhen14,bulg17pra,bulg17prl}.
It has been proved that  a generic propagating BIC on a periodic
structure with both the $C_2$ symmetry and the  reflection symmetry in
$z$ is robust against small structural perturbations  that preserve
these two symmetries \cite{zhen14,bulg17pra,yuan17_4,conrob}. 
It is also known that a BIC always turns to a resonant
state with a finite $Q$ factor, if the periodic structure is perturbed
by a generic perturbation breaking  the $C_2$ symmetry
\cite{kosh18,hu18,lijun20}.


In this Letter, we study how BICs depend on generic
  structural parameters. More precisely, in a generic parameter space,
  what is the dimension of the object formed by the BIC parameter values?
  It is clear that the answer depends on the number of
  opening radiation channels. For the simplest case of one opening
  radiation channel, we show that in the plane of two generic 
  parameters, the BICs exist as a curve, and in general, the BICs form
  a codimension-1 object in parameter space. This implies that 
on a periodic structure without the $C_2$ symmetry, it is possible to
find BICs by tuning one structural parameter. We
establish this result using a perturbation theory with 
two parameters.
Assuming a periodic structure has a BIC
with a frequency in the range for one opening radiation channel, we
consider a perturbed structure with two independent perturbations
where the amplitude of the first perturbation is fixed, and show that 
if the amplitude of the second perturbation is properly chosen,  then the
perturbed structure also has a BIC. The result requires 
some generic conditions on the original
BIC of the unperturbed structure and the perturbation profiles. 


To simplify the presentation, we consider a two-dimensional (2D)
dielectric structure that is invariant in $x$, periodic in $y$, and
symmetric in $z$. The dielectric function $\varepsilon({\bm r})$ for
${\bm r} = (y,z)$, satisfies   
\begin{equation}
  \label{refper}
  \varepsilon({\bm r}) = \varepsilon(y,-z)=\varepsilon(y+L, z)
\end{equation}
for all ${\bm r}$, where $L$ is the period. 
In addition, it is assumed
that the thickness of the periodic layer is $2d$ and the surrounding medium is
vacuum. Therefore, $\varepsilon({\bm r}) = 1$ for $|z| > d$. 
Notice that we do not assume a  reflection symmetry in $y$ which is
equivalent to the $C_2$ rotational symmetry for the 2D structure. For  
the $E$ polarization, the $x$ component of the electric
field, denoted as $u$, satisfies the following Helmholtz equation
\begin{equation}
  \label{helm}
  \partial_y^2 u + \partial_z^2 u + k^2 \varepsilon({\bm r}) u = 0, 
\end{equation}
where $k=\omega/c$ is the freespace wavenumber, $\omega$ is the
angular frequency and $c$ is the speed of light in vacuum. A BIC is a
Bloch mode solution of 
Eq.~(\ref{helm}) given as 
$u({\bm r}) = \phi({\bm r}) e^{ i \beta y}$, 
where $\phi$ is periodic in $y$ with period $L$ and decays to zero as 
$|z| \to \infty$, and $\beta$ is a real Bloch wavenumber satisfying
$|\beta | \le \pi/L$ and $ |\beta| < k$. 

First, we assume that a particular periodic
structure with a dielectric function $\varepsilon_*({\bm r})$
satisfying the conditions above has a
non-degenerate BIC   $u_* ({\bm r}) = \phi_*({\bm r}) e^{ i \beta_*
  y}$ with a frequency $\omega_*$ satisfying 
\begin{equation}
  \label{1channel}
|\beta_* | < k_* = \frac{\omega_*}{c}  < \frac{2\pi}{L} - |\beta_*|.  
\end{equation}
The above inequalities  ensure that there is only one opening radiation
channel. Due to the reflection symmetry in $z$, the BIC $u_*$ is either even in $z$
or odd in $z$. 

For the same $k_*$ and $\beta_*$, we
have diffraction problems for incident plane waves $\exp[ i (\beta_*
  y \pm \alpha_* z) ]$ where $\alpha_* = \sqrt{k_*^2 - \beta_*^2} > 0$, given
  in the media below and above the periodic layer. If the BIC is even
  in $z$, we can construct a diffraction solution that is also even in
  $z$.  We write the diffraction solution as 
  $v({\bm r})=  \psi({\bm r}) \exp (i \beta_*  y)$, where $\psi$ 
  is periodic in $y$ with period $L$, and require $v({\bm r})$ and the
  BIC be orthogonal, i.e., 
  \begin{equation}
    \label{ortho}
    \int_\Omega \varepsilon_*({\bm r}) \, \overline{v} \, u_* \, d{\bf
      r} = \int_\Omega \varepsilon_*({\bm r}) \, \overline{\psi} \,
    \phi_* \, d{\bm r} = 0, 
  \end{equation}
  where $\Omega$ is the domain for one period of the structure given
  by $|y| < L/2$ and $|z| < \infty$,   $\overline{v}$ is
  the complex conjugate of $v$, etc.  
This is possible, because the diffraction solution is not unique. If
$v({\bm r})$ is a diffraction solution, so is $v({\bm r}) + C u_*({\bm r})$
for any constant $C$. It is always possible to choose $C$ to redefine
$v$,  such that Eq.~(\ref{ortho}) is satisfied. Similarly, if the BIC is odd in $z$, there is an odd-in-$z$
diffraction solution $v({\bm r})$ satisfying the orthogonality
condition (\ref{ortho}). To analyze the parametric
  dependence of BICs, we need the following condition
\begin{equation}
  \label{a21}
  \int_\Omega \overline{v} \frac{\partial u_*}{\partial y} d{\bm r} =
  \int_\Omega  \overline{\psi} \left(  i \beta_*  \phi_* + 
    \frac{ \partial \phi_*}{\partial   y}  \right) \, d{\bm r} \ne 0.
\end{equation}
If the above is valid, we scale the diffraction solution or the BIC, such that 
\begin{equation}
  \label{a21p}
\int_\Omega \overline{\psi}  \left( \beta_* \phi_*  - i
  \frac{\partial \phi_*}{\partial y} \right)  d{\bm r}  
> 0. 
\end{equation}

If the periodic structure depends on parameters, the
  dielectric function may be written as 
  $\varepsilon = \varepsilon({\bm r}, {\bm p})$, where ${\bm p}$ is a
  vector of real parameters. Assuming a BIC exists for a particular
  parameter value ${\bm p}_*$, we study the existence of
  BICs for ${\bm p}$ near ${\bm p}_*$.
  Since $\varepsilon({\bf r}, {\bm p})$ can be approximated by
  $ \varepsilon({\bf r}, {\bm  p}_* ) + \nabla_{\bm p}
  \varepsilon({\bm r}, {\bm p}_*) \cdot ({\bm p} -{\bm p}_*)$,
  we consider, for the case of two parameters, a perturbed
periodic structure with a dielectric 
function given by 
\begin{equation}
  \label{peps}
\varepsilon({\bm r}) = \varepsilon_* ({\bm r}) + \delta G({\bm r}) + 
\gamma F({\bm r}), 
\end{equation}
where $\delta$ and $\gamma$ are components of ${\bm p}
  - {\bm p}_*$, $F$ and $G$ (the perturbation profiles) are partial
  derivatives of $\varepsilon({\bm r}, {\bm p})$ with respect to the
  components of ${\bm p}$.
 We assume both $F$ and $G$ are $O(1)$ real functions, are periodic in $y$
with period $L$ and symmetric in $z$, and vanish when $|z| > d$.
Thus, the perturbations preserve the periodicity in $y$ and the
reflection symmetry in $z$ and do not affect the surrounding
homogeneous media. In addition, we assume the perturbation profile $F$
satisfies 
\begin{equation}
  \label{condF}
\mbox{Im} \int_\Omega \overline{\psi}({\bm r}) F({\bm r}) \phi_* ({\bf
  r})  \, d{\bm r} \ne 0, 
\end{equation}
where $\phi_*$ and $\psi$ have their relative phase fixed by
condition (\ref{a21p}). 

Our main result can be stated as follows:  {\em If the periodic structure
with the dielectric function $\varepsilon_* ({\bm r})$ has a non-degenerate BIC $\{
\phi_*, k_*, \beta_* \}$ and an associated diffraction solution
$\psi({\bm r})$ satisfying conditions (\ref{1channel}) and
(\ref{a21p}), and the perturbation profile 
$F({\bm r})$ satisfies condition (\ref{condF}), then for any real
$\delta$ near $0$, there is a real $\gamma$ near 0, such that the
perturbed periodic structure with $\varepsilon({\bm r})$ given in
Eq.~(\ref{peps}) has a BIC 
$\{ \phi,  k, \beta \}$ with $\phi$ close to $\phi_*$, $k$ close to $k_*$ and 
$\beta$ close to $\beta_*$.}

The above proposition implies that the BICs  exist continuously as a
  curve in the plane of two generic parameters. It is clear that if
  there are $m$ generic parameters, the BICs should form a 
  codimension-1 object, i.e., an object with  dimension $m-1$, in the 
  $m$-dimensional parameter space.
To prove the above result, we expand the desired BIC 
and the parameter $\gamma$   as  power
series of $\delta$: 
\begin{eqnarray}
  \label{series1}
\phi &=& \phi_* + \phi_1 \delta + \phi_2 \delta^2 + \cdots \\
  \label{series2}
\beta &=& \beta_* + \beta_1 \delta + \beta_2 \delta^2 + \cdots\\
  \label{series3}
k &=& k_* + k_1 \delta + k_2 \delta^2 + \cdots\\
  \label{series4}
\gamma &=& \gamma_1 \delta + \gamma_2 \delta^2 + \cdots
\end{eqnarray}
and show that for each $j \ge 1$, $\beta_j$, $k_j$ and $\gamma_j$ can
be solved and they are real, $\phi_j$ can be solved and it decays to 0
as $|z| \to \infty$.
Briefly, we obtain  an inhomogeneous Helmholtz equation for
$\phi_j$ (see Supplemental Material \cite{supp}) 
and the following linear system for $\beta_j$, $k_j$ and 
$\gamma_j$: 
\begin{equation}
\label{3by3} 
 \left[ \begin{matrix} a_{11} & a_{12} & a_{13} \cr 
a_{21} & 0 & \mbox{Re}(a_{23}) \cr 
0 & 0 & \mbox{Im}(a_{23}) \end{matrix} \right] 
\left[ \begin{matrix} \beta_j\cr k_j\cr \gamma_j \end{matrix} \right] 
= \left[ \begin{matrix} b_{1j} \cr \mbox{Re}(b_{2j}) \cr
\mbox{Im}(b_{2j}) \end{matrix} \right], 
\end{equation}
where 
\begin{eqnarray}
  a_{11} &=& 2  \int_\Omega \overline{\phi}_* \left( \beta_* \phi_*
             -i \frac{ \partial \phi_*}{\partial y}  \right)  d{\bm r} \\
  a_{12} &=& -2 k_* \int_\Omega \varepsilon_* |\phi_*|^2 \, d{\bm r}  \\
a_{13} &=& -k_*^2 \int_\Omega F({\bm r}) |\phi_*|^2 \, d{\bm r}\\
  a_{21} &=& 2   \int_\Omega\overline{\psi} \left( \beta_* \phi_*   
             -i  \frac{ \partial \phi_*}{\partial y} \right) d{\bm r} \\
a_{23} &=& -k_*^2 \int_\Omega F({\bm r}) \overline{\psi} \phi_* \, d{\bm r}, 
\end{eqnarray}
$b_{1j}$ and $b_{2j}$ can also be explicitly written down (see
Supplemental  Material \cite{supp}). 
Using integration by parts, it is easy to show that $a_{11}$ is
real. From condition (\ref{a21p}),  we see that $a_{21}$ is 
also real. Therefore, all entries in the $3 \times 3$ coefficient
matrix of Eq.~(\ref{3by3}) are real. The second column the
coefficient matrix is related to 
\begin{equation}
a_{22} =  -2 k_* \int_\Omega \varepsilon_*  \overline{\psi} \phi_* \, d{\bm 
  r},   
\end{equation}
and it is zero by the orthogonality assumption (\ref{ortho}). 
Since $a_{12}$, $a_{21}$ and
$\mbox{Im}(a_{23})$ [by condition (\ref{condF})], are all nonzero, the
coefficient matrix is invertible.
Moreover, we can prove that $b_{1j}$ is real for all $j\ge 1$ (see
Supplementary Material \cite{supp}),  therefore, for each $j\ge 1$,
$\beta_j$, $k_j$ and $\gamma_j$ can be solved and they are real. 
Meanwhile, we can show that the governing equation for $\phi_j$ has a 
solution with the same parity (even or odd in $z$) as $\phi_*$, and it 
decays to zero as $|z| \to \infty$ (see Supplemental Material
\cite{supp}). 

Notice that $G({\bm r})$ has no additional condition besides those
specified below Eq.~(\ref{peps}). If $G({\bm r}) \equiv  0$, the above
procedure does not give a new BIC, namely, we only get $\gamma=0$
and $u = u_*$. The same is true if $F({\bm r})$ and $G({\bm r})$ are
proportional to each other. For example, if $F({\bm r}) = G({\bm r})$, then 
$\gamma = - \delta$ and $u = u_*$. 

If the original BIC is a standing wave ($\beta_* =0$),  then the BIC
on the perturbed structure is also a standing wave. Since $\beta_* =0$,
and the real and imaginary 
parts of a standing wave $u_*$ are also standing waves. Therefore, we
can assume $u_* ({\bm r})=\phi_* ({\bm r})$ is a real function. The
diffraction solution $v({\bm r})$ contains normally incident 
and reflected plane waves below and above the periodic layer, and it can be
scaled to be a pure imaginary function. Thus,  $a_{23}$ is pure
imaginary. In addition, we can show that for each $j \ge 1$, 
$b_{2j}$ is pure imaginary (see Supplemental Material
\cite{supp}). Therefore, the second equation in  (\ref{3by3}) is
$a_{21} \beta_j = 0$ and it gives $\beta_j = 0$ for 
all $j \ge 1$. This result relies on the non-degeneracy of the
standing wave and is consistent with the reciprocity principle. If the
perturbed structure has a BIC with Bloch wavenumber $\beta \ne
0$, it must have another one with 
Bloch wavenumber $-\beta$, and both of them are perturbations
of the original standing wave $u_*$. However, the power series
(\ref{series1})-(\ref{series4}) give only one solution. This implies
that the BIC on the perturbed structure must also be a standing wave.

If the original periodic structure has an additional reflection symmetry in
$y$ and $G({\bm r})$ is also symmetric in $y$, then according to our
previous work \cite{yuan17_4}, a  perturbed 
structure with a dielectric function $\varepsilon_* ({\bm r}) + \delta
G({\bm r})$ should have a 
BIC near the original one on the unperturbed structure. Our theory 
developed above covers this result as a special case. If $F({\bm r})$ is not
symmetric in $y$, we can show  that $\gamma_j = 0$ for all $j \ge
1$ (see Supplemental Material \cite{supp}). Therefore, a BIC exists on
the perturbed structure  
with only one perturbation $ \delta G({\bm r})$. If $F({\bm r})$ is
also symmetric in $y$, then condition 
(\ref{condF}) is not satisfied, thus, $\gamma$ cannot be
determined as a function of $\delta$. In fact, for any small
$\gamma$, $\delta G({\bm r}) + \gamma F({\bm r})$ is a perturbation
preserving the reflection symmetry in $y$, and there should be a  BIC on
the perturbed structure. 

\begin{figure}[h]
  \centering
 \includegraphics[scale=0.7]{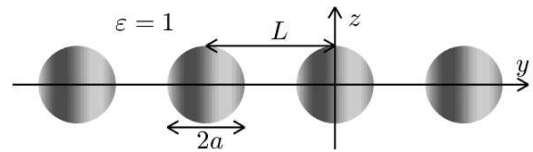}
  \caption{A periodic array of circular cylinders with radius $a$ and
    a $y$-dependent dielectric function. The array is symmetric in
    $z$ and asymmetric in $y$.}
  \label{1array}
\end{figure}

To validate the above theory, we consider a periodic array of circular
cylinders surrounded by vacuum as shown in
Fig.~\ref{1array}. 
The radius of the cylinders is $a=0.3L$, where $L$ is the period in
the $y$ direction. The dielectric function $\varepsilon({\bm r)}$ of the structure is 
given by Eq.~(\ref{peps}) with $\varepsilon_*({\bm r}) = 10$, 
\[
F({\bm r})  =  \sin \left( \dfrac{\pi y}{2 a} + \dfrac{\pi}{4}\right), \
G({\bm r}) = \sin \left( \dfrac{\pi y}{a}\right), 
\]
if ${\bm r}$ is inside the cylinders, i.e., for an integer $m$, $y$ and
$z$ satisfy $(y-m L)^2 + z^2 < a^2$, and 
$\varepsilon_* ({\bm r}) = 1$ and $F({\bm r}) = G({\bm r}) =
0$ if ${\bm r}$ is outside the cylinders.  For nonzero $\delta$ and
$\gamma$, the periodic array is symmetric in $z$ and asymmetric in
$y$.  For $\delta=\gamma=0$, the periodic array is symmetric in $y$,
and has a standing wave
with freespace wavenumber $k_* = 0.4414 (2\pi/L)$ and a propagating BIC
with $k_* =0.6173 (2\pi/L)$ and Bloch wavenumber $\beta_* = 0.2206
(2\pi/L)$. The standing wave is anti-symmetric in $y$, i.e., it is a
symmetry-protected BIC. 

\begin{figure}[h]
  \centering
  \includegraphics[scale=0.6]{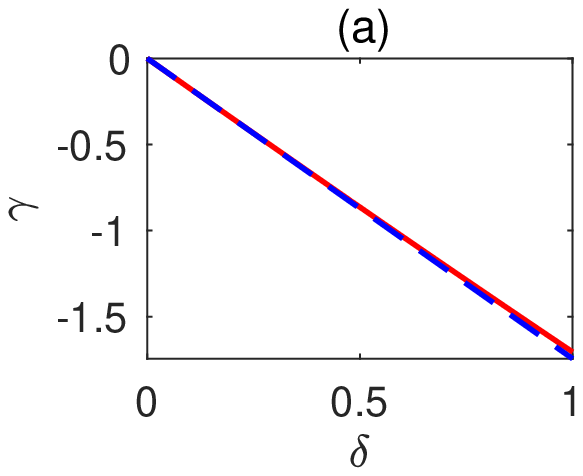}
  \includegraphics[scale=0.6]{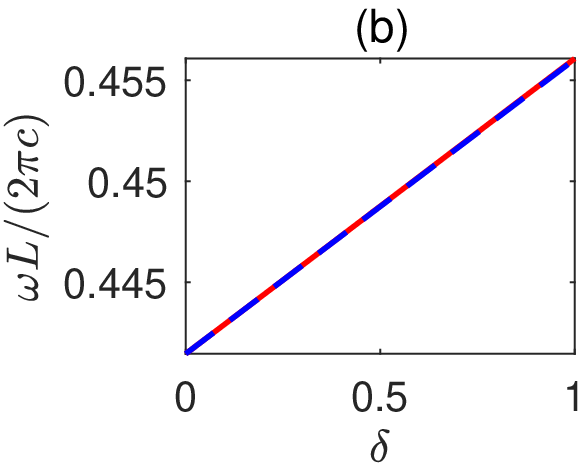}
  \includegraphics[scale=0.6]{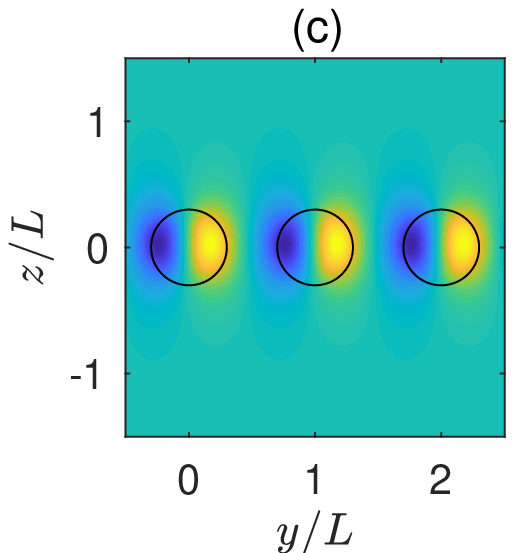}
  \includegraphics[scale=0.6]{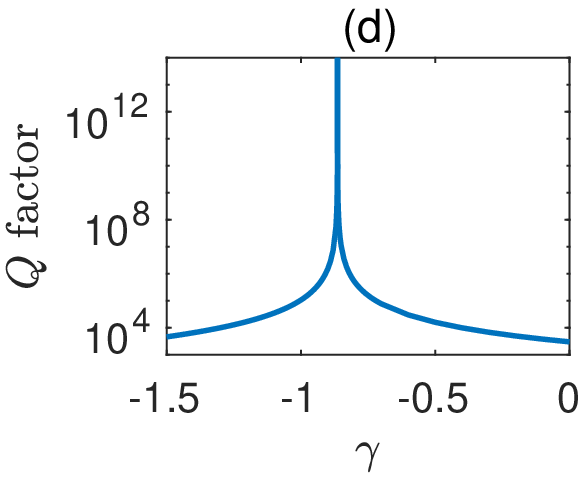}
  \caption{Standing wave on a periodic array without a reflection
    symmetry in $y$:  (a) $\gamma$ as a function of $\delta$ for the
    existence of the standing wave;  (b) frequency of  the standing
    wave as a function of $\delta$; (c) wave field pattern of the standing wave for
     $\delta = 0.5$ and $\gamma = -0.863673$; (d) $Q$ factors of the
     resonant states for 
     $\beta=0$, $\delta=0.5$ and different values of $\gamma$. }
  \label{stand}
\end{figure}

For a proper  $\gamma$ depending on $\delta$, the BICs exist
continuously with respect to $\delta$. Using an efficient numerical
method, we calculated 
the BICs for $0 \le \delta \le 1$. As predicted by our theory, the
standing wave for $\delta=\gamma=0$ continues as a standing wave for
$\delta \ne 0$. In
Figs.~\ref{stand}(a) and \ref{stand}(b), 
the solid red curves depict $\gamma$ and freespace wavenumber
$k$ as functions of $\delta$ for 
the standing wave. Based on the standing wave for $\delta=\gamma=0$,
we found the first order perturbation terms $k_1 = 0.0146 (2\pi/L)$ and
$\gamma_1=-1.7491$. The first order perturbation results, i.e. $\gamma
\approx 
\gamma_1 \delta$ and $ k \approx k_*  + k_1 \delta$ are shown in
Figs.~\ref{stand}(a) and \ref{stand}(b) as the blue dashed lines, and
they are quite accurate even when $\delta$ is not small. For
$\delta=0.5$, the standing wave is obtained at $\gamma = -0.863673$
and it is shown in Fig.~\ref{stand}(c). For other values of $\gamma$,
the periodic structure can only have resonant states with complex
frequencies. In Fig.~\ref{stand}(d), we show the $Q$ factors of the 
resonant states as a function of $\gamma$ for $\delta = 0.5$ and $\beta =
0$, where  the infinite peak corresponds to the standing wave. 

\begin{figure}[h]
  \centering
  \includegraphics[scale=0.6]{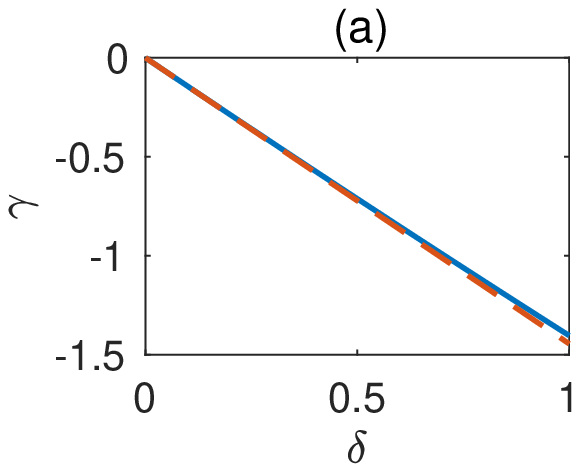}
  \includegraphics[scale=0.6]{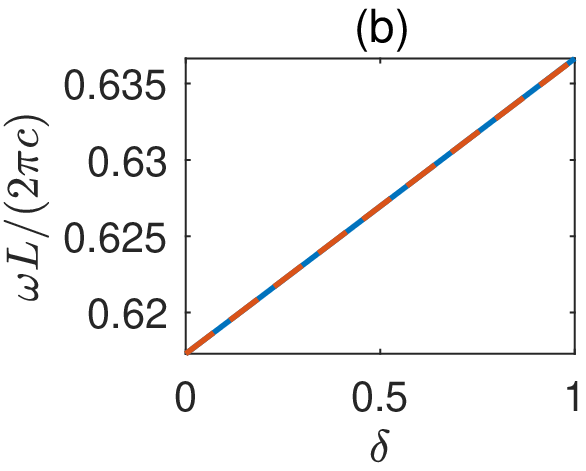}\\
  \includegraphics[scale=0.6]{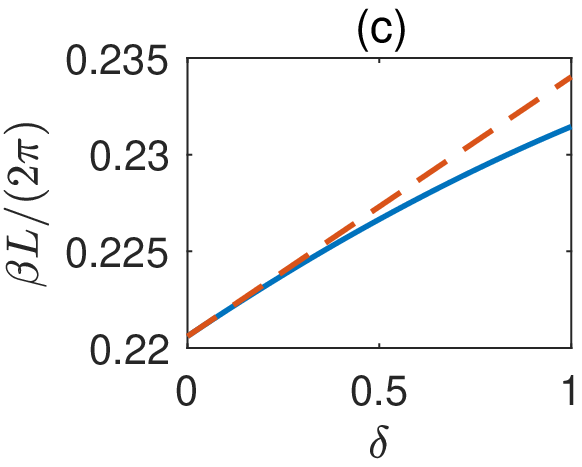}
  \includegraphics[scale=0.56]{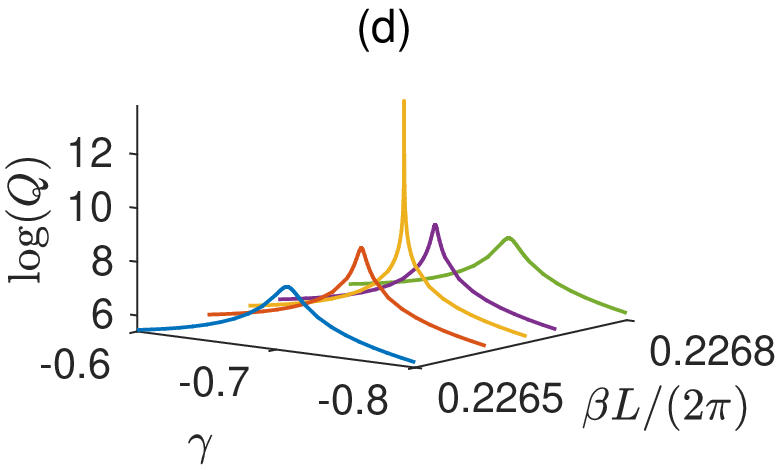}
  \includegraphics[scale=0.55]{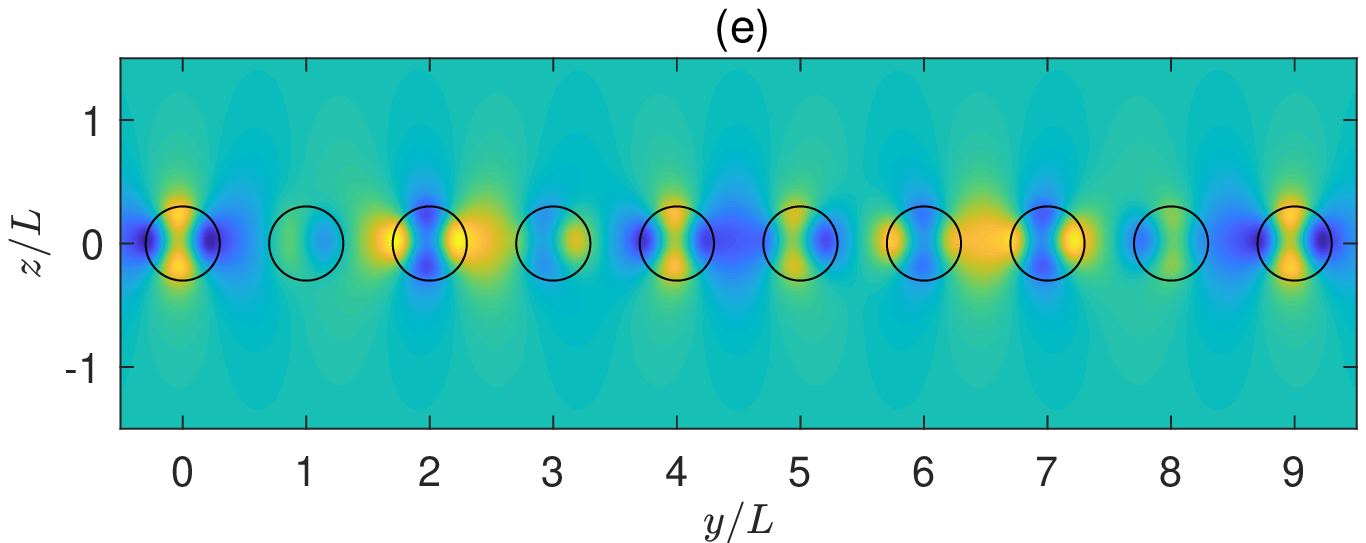}
  \caption{Propagating BIC on a periodic array without  a reflection
    symmetry in $y$: (a) $\gamma$ as a function of $\delta$ for the
    existence of the BIC; (b) frequency of the propagating  BIC; (c)
    Bloch wavenumber of the propagating BIC; (d) $Q$ factors of the
    resonant states for $\delta=0.5$ and different values of $\beta$
    and $\gamma$; (e) wave field pattern, i.e., $ \mbox{Re}[u({\bm r})]$, of
    the propagating BIC for $\delta = 0.5$ and $\gamma = -0.711932$.}
  \label{pbic}
\end{figure}

The propagating BIC for $\delta=\gamma=0$ can also be easily extended
to the case with $\delta \ne 0$. In 
Figs.~\ref{pbic}(a),  \ref{pbic}(b) and \ref{pbic}(c), 
we show $\gamma$, $k$ and $\beta$ as functions of
$\delta$ for the 
propagating BIC, respectively. The first order
perturbation terms are found as $\gamma_1 = -1.4445$, $k_1 = 0.0193 (2\pi/L)$, and
$\beta_1 = 0.0134 (2\pi/L)$. In Figs.~\ref{pbic}(a), \ref{pbic}(b) and \ref{pbic}(c), the first order perturbation results
are shown as the red dashed lines. For $\delta = 0.5$ and 
$\gamma =  -0.711932$, the periodic structure supports a propagating
BIC with $ k = 0.626957 (2\pi/L)$ and 
$\beta = 0.226658 (2\pi/L)$. 
Its field pattern is shown in
Fig.~\ref{pbic}(e).  
As shown in Fig.~\ref{pbic}(d), the periodic structure has 
resonant states with finite $Q$ factors for nearby values of $\gamma$
and $\beta$.

In summary, we show analytically that  if there is
  only one radiation channel, typical BICs on periodic structures form
  codimension-1 objects in the space of generic structural
  parameters, and consequently,
on periodic structures without the 
$C_2$ rotational symmetry, it is possible to find BICs by tuning one structural
parameter. 
The result is established by analyzing a perturbation with
  two independent terms.
Some generic conditions
are needed for the periodic structure, the perturbations and the
original BIC. The reflection symmetry in $z$ and condition (\ref{1channel}) are 
imposed such that there is only one independent radiation channel. The 
BIC must be non-degenerate and must satisfy condition (\ref{a21}), and
the perturbation profile $F({\bm r})$ must satisfy condition
(\ref{condF}). Our theory and numerical results are presented for 2D periodic
structures with a single periodic direction. We are currently
completing an extension for three-dimensional bi-periodic
structures. It may be possible to develop similar theories for BICs
on other structures, such as rotationally-symmetric periodic
structures \cite{bulg17prl,sadbel19} and waveguides with
lateral leaky channels \cite{byk20}.

The authors acknowledge support from the Natural Science Foundation of
Chongqing, China, under Grant No. cstc2019jcyj- msxmX0717,  and the
Research Grants Council of Hong Kong Special Administrative Region,
China, under Grant No. CityU 11304117.

\onecolumngrid 
\newpage
\begin{center}
{\bf \large Supplemental material for ``Bound States in the Continuum on
  Periodic Structures without $C_2$ Rotational Symmetry''}
\end{center} 
\renewcommand{\theequation}{S\arabic{equation}}
\renewcommand{\thefigure}{S\arabic{figure}}
\renewcommand{\bibnumfmt}[1]{[S#1]}
\renewcommand{\citenumfont}[1]{S#1}

\section{S1: Perturbation process}

On a 2D periodic structure that is invariant in $x$, periodic in $y$
with period $L$, and symmetric in $z$, a Bloch mode for the $E$
polarization is given as $u({\bf r}) = \phi({\bf r}) e^{ i \beta y}$ for ${\bf
  r}=(y,z)$, where $\phi$ satisfies 
\begin{equation}
\label{eq:helm2} \frac{\partial^2 \phi}{\partial z^2} +   \frac{\partial^2
  \phi}{\partial y^2} + 2 i \beta \frac{\partial \phi}{\partial y} + [
k^2 \varepsilon({\bf r})  - \beta^2  ] \phi = 0. 
\end{equation}
The dielectric function $\varepsilon({\bf r})$ of the
periodic structure is assumed to be 
$\varepsilon({\bf r}) = \varepsilon_* ({\bf r})  +
\gamma F({\bf r}) + \delta G({\bf r})$, 
where $\varepsilon_*({\bf r})$ is the dielectric function of the original
unperturbed structure, $F$ and $G$ are perturbation profiles, $\delta$
and $\gamma$ are small real parameters.
We assume the unperturbed structure supports a BIC $u_*({\bf r}) =
\phi_*({\bf r}) e^{ i \beta_* y}$ with frequency $\omega_*=k_* c$. Since
the unperturbed structure also has a reflection symmetry in $z$, we
can construct a diffraction solution $v({\bf r}) = \psi({\bf r})e^{ i \beta_*
  y}$ with the same even or odd parity in $z$ as the BIC $u_*({\bf
  r})$.  The diffraction solution is not unique, we can choose
$v$ such that 
\begin{equation}
  \label{ortho}
  \int_\Omega \varepsilon_* \overline{\psi} \phi_* d{\bf r} = 0,
\end{equation}
where $\Omega$ is given by $|y| < L/2$ and $|z| < \infty$. Moreover,
we assume 
\begin{equation}
     \int_\Omega \overline{v} \frac{\partial u_*}{\partial y} d{\bf r}
  \ne 0. 
\end{equation}
If the above is true, we scale the diffraction solution, such that
\begin{equation}
\label{a21p}
  \int_\Omega \overline{\psi} \left[ \beta_* \phi_* - i \frac{\partial
      \phi_*}{\partial y} \right] d{\bf r}      > 0.
   \end{equation}
Finally, the  perturbation profile $F$ is required to satisfy 
\begin{equation}
  \label{condF}
  \mbox{Im}   \int_\Omega F \overline{v} u_* d{\bf r}
  =   \mbox{Im}   \int_\Omega F \overline{\psi} \phi_* d{\bf r}
  \ne 0.  
\end{equation}

For  a given small $\delta$, we try to determine $\gamma$ such that the
Bloch mode $u({\bf r})$ is a BIC. Expanding $\phi$, $\beta$, $k$ and
$\gamma$ in power series of $\delta$, and comparing coefficients of
$\delta^j$ for $j \ge 1$, we obtain 
\begin{equation}
\label{phij} \mathcal{L} \phi_j = B_1 \beta_j + B_2 k_j + B_3 \gamma_j - C_j,
\end{equation}
where 
\begin{eqnarray*}
  \mathcal{L} &=& \partial_z^2 + \partial_y^2 + 2 i \beta_* \partial_y
   + k_*^2 \varepsilon_* - \beta_*^2\\
  B_1  &=& 2 \beta_* \phi_* - 2 i \frac{\partial \phi_*}{\partial y} \\
  B_2  &=& -2 k_* \varepsilon_* \phi_* \\
  B_3  &=& - k_*^2 F  \phi_* \\
  C_j  &=&  V_j \phi_* + \sum\limits_{n=1}^{j-1} W_n \phi_{j-n} + 2 i
           \sum\limits_{n=1}^{j-1} \beta_n \frac{ \partial
           \phi_{j-n}}{\partial y} \\
 V_j &=& \sum\limits_{m=1}^{j-1} (k_m k_{j-m} \varepsilon_* - \beta_m
         \beta_{n-m})  + G \sum\limits_{m=0}^{j-1} k_m k_{j-1-m} +  F
         \sum\limits_{m=1}^{j-1}  \sum\limits_{l=0}^{m} k_l k_{m-l}
         \gamma_{j-m} \\
  W_n &=& V_n + 2 k_* k_n \epsilon_*  - 2 \beta_* \beta_n + k_*^2 \gamma_n F, 
          \qquad 1 \leq n \leq j-1.
\end{eqnarray*}
In the above, we set $\beta_0 = \beta_*$,  $k_0 = k_*$, $\gamma_0 =
0$,  and $\phi_0 = \phi_*$ in the sum terms. Notice that $C_j$ only
involves $\beta_m$,  $k_m$, $\gamma_m$ and $\phi_m$ for $m \leq
j-1$. More specifically, 
\begin{eqnarray*}
&& C_1 = V_1 \phi_* = k_*^2 G \phi_* \\
  && C_2 = V_2 \phi_* +  W_1 \phi_1 + 2 i  \beta_1 \partial_y \phi_1 \\
     &&= (k_1^2 \varepsilon_* - \beta_1^2 + 2 k_* k_1 G + 2 k_* k_1
  \gamma_1 F) \phi_* + (2k_* k_1 \varepsilon_* - 2 \beta_* \beta_1 + k_*^2 G +
  k_*^2 \gamma_1 F) \phi_1 +2 i  \beta_1 \partial_y \phi_1.
\end{eqnarray*}

If Eq.~(\ref{phij}) has a solution $\phi_j$ that decays to 0 as $|z|
\to \infty$, we can multiple $\overline{\phi}_*$ and $\overline{\psi}$
to both sides and integrate on $\Omega$. This gives rise to
\begin{equation}
  \label{lin_sys}
\left[ \begin{matrix} a_{11} & a_{12} & a_{13} \\ a_{21} & a_{22} &
    a_{23}   \end{matrix} \right]
   \left[ \begin{matrix} \beta_j \\ k_j \\ \gamma_j  \end{matrix}
   \right] =
\left[ \begin{matrix} b_{1j} \\ b_{2j} \end{matrix} \right],    
\end{equation}
where 
\begin{eqnarray*}
&&  a_{1m} = \int_{\Omega} \overline{\phi}_* B_m d \mathbf{r}, 
  \quad a_{2m} = \int_{\Omega} \overline{\psi} B_m d \mathbf{r}, 
   \qquad m = 1, 2, 3 \\
&& b_{1j} = \int_{\Omega} \overline{\phi}_* C_j d \mathbf{r}, \quad
   b_{2j} = \int_{\Omega} \overline{\psi}  C_j d \mathbf{r}.
\end{eqnarray*}
It is easy to check that $a_{11}$, $a_{12}$ and $a_{13}$ are all real.
From conditions (\ref{ortho}) and (\ref{a21p}), we obtain  $a_{22} = 0$
and $a_{21} > 0$.  The linear system (\ref{lin_sys}) can be written as 
\begin{equation}
  \label{lin_sys2}
  \left[ \begin{matrix} a_{11} & a_{12} & a_{13} \\
      a_{21}& 0  & \mbox{Re}(a_{23}) \\ 
      0 & 0  & \mbox{Im}(a_{23}) \end{matrix} \right]
\left[ \begin{matrix} 
    \beta_j  \cr k_j  \cr \gamma_j
  \end{matrix} \right]
 = \left[ \begin{matrix}
 b_{1j} \cr
 \mbox{Re}(b_{2j}) \cr
\mbox{Im}(b_{2j}) \end{matrix} \right].
\end{equation}
Since $F$ must satisfy condition (\ref{condF}), 
$\mbox{Im}(a_{23}) \ne 0$. Therefore, the $3 \times 3$ coefficient
matrix above  is invertible. To have real solutions for 
$\beta_j$, $k_j$ and $\gamma_j$, we need to show that $b_{1j}$ is real
for all $j \geq 1$.

For $j=1$, $C_1 = k_*^2 G \phi_*$. Since $G$ is real, $b_{11}$ is
real. Meanwhile, since $C_1$ decays exponentially as $|z| \to \infty$,
$b_{21}$ is well defined.  Therefore, we can find real $\beta_1$, $k_1$ and
$\gamma_1$. After that, we can solve $\phi_1$ from Eq.~(\ref{phij}) for
$j=1$. The solution $\phi_1$ decays to zero  as $|z| \to \infty$ and has the same
even/odd parity in $z$  as $\phi_*$. 

For $j=2$, $C_2 = V_2 \phi_*  +  W_1 \phi_1 + 2 i  \beta_1 \partial_y
\phi_1$. It decays exponentially as $|z| \to \infty$ and has the same
even/odd parity as $\phi_*$. Thus, $b_{22}$ is well defined.
Since $V_2$ is real and 
\[
  b_{12} = \int_{\Omega} V_2 |\phi_*|^2 d \mathbf{r} + \int_{\Omega}
  \left( W_1 \phi_1 + 2i \beta_1 \partial_y \phi_1 \right)
  \overline{\phi}_*  d \mathbf{r},
\]
it is only necessary to show the second term on the right hand side above is
real.  Multiplying the complex conjugate of Eq.~(\ref{phij}) for $j=1$
by $\phi_1$ and integrating  on $\Omega$, we have 
\begin{eqnarray*}
  \int_{\Omega} \phi_1 \overline{\mathcal{L} \phi_1} d \mathbf{r}&= &
  - \int_{\Omega} ( 2 k_* k_1 \varepsilon_* - 2 \beta_* \beta_1 +
     k_*^2 \gamma_1 F + V_1) \overline{ \phi}_* \phi_1 d \mathbf{r}
    + 2i \beta_1 \int_{\Omega}  \phi_1 \partial_y \overline{\phi}_* d \mathbf{r} \\
 & = & - \int_{\Omega} ( W_1 \phi_1 + 2 i \beta_1 \partial_y \phi_1)   \overline{\phi}_*  d \mathbf{r}. 
\end{eqnarray*}
It is easy to check that $\int_{\Omega} \phi_1 \overline{\mathcal{L}
  \phi_1} d \mathbf{r}$ is real, thus $b_{12}$ is real. Solving
Eq.~(\ref{lin_sys2}) for $j=2$, we obtain real $\beta_2$, $k_2$ and
$\gamma_2$. After that, $\phi_2$ can be solved from Eq.~(\ref{phij})
for $j=2$, and it decays to zero as $|z| \to \infty$ and has the same 
even/odd parity as $\phi_*$. 

For $j > 2$, we  assume  for any $n$ satisfying  $1 \leq n \leq  j-1$, $\beta_n$, $k_n$ and $\gamma_n$ are
 real,  $\phi_n \to 0$ as $|z| \to \infty$, and $\phi_n$ has  the same
 even/odd parity as $\phi_*$. This implies that 
 $V_j$ and $W_n$ for $1
 \leq n \leq j-1$ are real.
 Since 
 \[
   b_{1j} = \int_{\Omega} \overline{\phi}_* C_j d \mathbf{r} =
   \int_{\Omega} V_j |\phi_*|^2 d \mathbf{r} + \sum\limits_{n=1}^{j-1}
   \int_{\Omega} \left( W_n  \phi_{j-n}  + 2 i \beta_n  \partial_y
     \phi_{j-n} \right) \overline{\phi}_* d \mathbf{r},
 \]
we only need to show the last  term on the right hand side of above is
real. Using the relations between $W_n$ and $V_n$ for $1 \leq n \leq
j-1$, we obtain 
\begin{eqnarray*}
  \sum\limits_{n=1}^{j-1} \int_{\Omega} \phi_{j-n} \overline{\mathcal{L} \phi_n} d \mathbf{r} & = &  -\sum_{n=1}^{j-1} \int_{\Omega} \left( W_n  \phi_{j-n}  + 2 i \beta_n  \partial_y \phi_{j-n} \right) \overline{\phi}_*  d \mathbf{r}   \\
& - & \sum\limits_{n=1}^{j-2} \int_{\Omega} \left( \sum\limits_{m=n+1}^{j-1} \overline{\phi}_{m-n} \phi_{j-m} \right) W_n d \mathbf{r} + 2i \sum\limits_{n=1}^{j-2}  \beta_n  \int_{\Omega} \left( \sum\limits_{m=n+1}^{j-1}  \phi_{j-m} \partial_y \overline{\phi}_{m-n} \right)d \mathbf{r} .
\end{eqnarray*}
It is easy to verify that $\sum\limits_{n=1}^{j-1} \int_{\Omega}
\phi_{j-n} \overline{\mathcal{L} \phi_n} d \mathbf{r} $ and
$\sum\limits_{m=n+1}^{j-1} \overline{\phi}_{m-n} \phi_{j-m}$
for $1 \leq n \leq j-1$ are real,  
 and $\sum\limits_{m=n+1}^{j-1}  \phi_{j-m} \partial_y
\overline{\phi}_{m-n}$ for $1 \leq n \leq j-1$ are pure imaginary.
Therefore, the first term on the right hand side of above is real, and
$b_{1j}$ is real.  

\section{S2: Special cases}


If the BIC of the unperturbed structure is a standing wave, i.e.,
$\beta_* = 0$,  then we can choose $\phi_*$  as a real function and
$\psi$ as a pure imaginary function. In that case, 
$a_{11} = 0$, 
$a_{21}$ is real, $a_{23}$ and $b_{2j}$ must be pure imaginary.  The
second equation of linear system Eq.~(\ref{lin_sys2}) gives 
$a_{21} \beta_j  =  0$ for all $j \ge 1$. Therefore, $\beta_j = 0$ for
all $j \ge 1$.  


If the unperturbed structure has a reflection symmetry in $y$, i.e.,
$\varepsilon_*$ satisfies 
$\varepsilon_*({\bf r}) = \varepsilon_*(-y,z)$ for all ${\bf r}$, 
then, as shown in Ref.~\cite{yuan17_4}, we can scale the BIC and the
diffraction solution such that  
$\phi_*$ and $\psi$ are ${\cal PT}$-symmetric in $y$, i.e., 
\begin{equation}
  \label{PTiny}
\phi_*({\bf r}) = \overline{\phi}_*(-y,z), \quad 
\psi({\bf r}) = \overline{\psi}(-y,z)
\end{equation}
for all ${\bf r}$. 
We still assume conditions (\ref{ortho}) and (\ref{a21p}) are satisfied,
since they are consistent with the ${\cal PT}$-symmetry
scaling. If $F$ is not symmetric in $y$, then $a_{23} = -k_*^2
\int_\Omega F \psi \phi_* d{\bf r}$ is in general complex, and we can
assume condition (\ref{condF}) is satisfied. 
Therefore, the coefficient matrix in 
(\ref{lin_sys2}) is real and invertible. 
If $G$ is symmetry in $y$, then 
$b_{21} = k_*^2 \int_\Omega G \overline{\psi} \phi_* d{\bf r}$ is
real. Thus, linear system (\ref{lin_sys2}) gives 
a real $\beta_1$, a real $k_1$, and $\gamma_1=0$. 
The right hand side of the governing equation
for $\phi_1$ does not involve $F$ and is $PT$-symmetric. This implies
that $\phi_1$ and $C_2$ are ${\cal PT}$-symmetric. Repeating 
this process, we obtain $\gamma_j = 0$ for all $j \ge 1$.

\end{document}